# Nanosecond Phase Transition Dynamics in Compressively Strained Epitaxial BiFeO$_3$

*Margaret P. Cosgriff, Pice Chen, Sung Su Lee, Hyeon Jun Lee, Lukasz Kuna, Krishna C. Pitike, Lydie Louis, William D. Parker, Hiroo Tajiri, Serge M. Nakhmanson, Ji Young Jo, Zuhuang Chen, Lang Chen, and Paul G. Evans\**

Dr. M. P. Cosgriff, Dr. P. Chen,[+] Prof. P. G. Evans,
Department of Materials Science and Engineering and Materials Science Program, University of Wisconsin-Madison, Madison, Wisconsin 53706, USA
E-mail: pgevans@wisc.edu

S. S. Lee, H. J. Lee, Prof. J. Y. Jo
School of Materials Science and Engineering, Gwangju Institute of Science and Technology, Gwangju 500712, South Korea

L. Kuna,
Department of Physics, University of Connecticut, Storrs, Connecticut 06269, USA

K. C. Pitike, Dr. L. Louis,
Department of Materials Science and Engineering and Institute of Materials Science, University of Connecticut, Storrs, Connecticut 06269, USA

Prof. S. M. Nakhmanson,
Department of Physics, Department of Materials Science and Engineering, and Institute of Materials Science, University of Connecticut, Storrs, Connecticut 06269, USA

Dr. W. D. Parker,
Argonne Leadership Computing Facility, Argonne National Laboratory, Argonne, Illinois 60439, USA and School of Chemical Engineering, Purdue University, West Lafayette, Indiana 47907, USA

Dr. H. Tajiri,
Japan Synchrotron Radiation Research Institute, SPring-8, Hyogo 6795198, Japan

Dr. Z. Chen,
Department of Materials Science and Engineering, University of California, Berkeley, California 94720, USA

Prof. L. Chen,
Department of Physics, South University of Science and Technology of China, Shenzhen, China 518055

[+] Present address: Department of Materials Science and Engineering, Northwestern University, Evanston, Illinois 60208, USA

A highly strained $BiFeO_3$ (BFO) thin film is transformed between phases with distinct structures and properties by nanosecond-duration applied electric field pulses. Time-resolved synchrotron x-ray microdiffraction shows that the steady-state transformation between phases is accompanied by a dynamical component that is reversed upon the removal of the field. Steady-state measurements reveal that approximately 20% of the volume of a BFO thin film grown on a $LaAlO_3$ substrate can be reproducibly transformed between rhombohedral-like and tetragonal-like phases by electric field pulses with magnitudes up to 2 MV/cm. A transient component, in which the transformation is reversed following the end of the electric field pulse, can transform a similar fraction of the BFO layer and occurs rapidly time scale limited by the charging time constant of the thin film capacitor. The piezoelectric expansion of the tetragonal-like phase leads to a strain of up to 0.1%, with a lower limit of 10 pm/V for the piezoelectric coefficient of this phase. Density functional theory calculations provide insight into the mechanism of the phase transformation showing that imparting a transient strain of this magnitude favors a transformation from rhombohedral-like to tetragonal-like phase.

# 1. Introduction

The functional response of materials to external fields can be dramatically enhanced by exploiting the sensitivity to small perturbations at magnetic, structural, and electronic phase transitions.[1] In the case of complex oxides, enhanced piezoelectricity and magnetoresistance have resulted from the discovery of materials that are near structural or magnetic phase boundaries.[2,3] Typically, adjusting the composition of a material is what places it near a phase transition, as is the case for colossal magnetoresistive oxides and relaxor ferroelectrics.[2,3] The multiferroic complex oxide $BiFeO_3$ (BFO), can however be driven into the vicinity of a phase boundary by varying the degree of strain imparted in a BFO thin film layer through the elastic constraint that heteroepitaxy imposes.[4] At this boundary there is a complex coexistence of structural phases that can be alternatively described using monoclinic or triclinic symmetries.[5-7] The transformation of BFO from rhombohedral-like to tetragonal-like symmetry at the phase boundary can be driven by an applied electric field and is accompanied by a very large lattice expansion along the out-of-plane direction in thin film samples.[8] The effective piezoelectric expansion due to the structural phase transformation reaches several percent.[8]

Piezoelectric strain-electric field hysteresis loop studies show that electric fields with either sign can induce the transformation into the tetragonal-like phase.[8] Effects associated with polarization imprint can make the thresholds for the transformation asymmetric with respect to the applied field, such that fields with a magnitude on the order of +/-1.50 MV cm$^{-1}$ can drive a compressively strained BFO film into the tetragonal-like phase after a positive pulse and back to the rhombohedral-like phase after a negative pulse.[7] The extent of this transformation depends on the magnitude, duration, and sign of the applied field. The dynamics of the transformation, its structural mechanism, and its reproducibility after large numbers of repeated field pulses have remained unclear.

Here we show that the electric-field-induced structural transformation can occur on nanosecond timescales limited, in this case, only by the charging time constant of the thin film capacitor. Synchrotron x-ray diffraction studies show that the film transforms between phases within nanoseconds and that there are components of the transformation that are reversed when the applied field is removed. The transient transformation is reproducible over the course of more than $10^8$ repetitions of the applied electric field pulses. Free-energy density functional theory (DFT) calculations accurately predict the structural mechanism of the transition. Taken together, these results promise to enable nanosecond-scale control of the crystallographic phase of materials, and hence functional properties, using applied electric fields.

## 2. Experiment

Time-resolved x-ray diffraction measurements were performed at station BL13XU of the SPring-8 synchrotron light source, using x-rays with photon energy of 12.4 keV focused to a 2 μm full width at half maximum spot on the sample using a Si refractive lens.[9] Diffracted x-rays were detected with a gated two-dimensional pixel array detector (Pilatus 100K, Dectris Ltd.) synchronized to the applied voltage pulses to enable time-resolved measurements of the response of the BFO film to an applied electric field.[10] An epitaxial BFO film with a thickness of 70 nm was deposited at a temperature of 700 °C and pressure of 100 mTorr onto a LaNiO3 (LNO) bottom electrode on an LaAlO3 (LAO) substrate. Gold top electrodes with 55 μm diameter were deposited on the BFO surface through a shadow mask to form thin-film capacitors with a capacitance of 9 pF. The LNO bottom electrode had a sheet resistance of 3.3 kΩ, which is approximately the resistance associated with charging the BFO thin film capacitors, thus yielding charging time constants of 30 ns. Voltage pulses were applied across the thin film capacitors using a probe tip contacting the top electrode, with the bottom electrode grounded. Voltage pulses had either square or triangular waveforms with positive or negative magnitudes of up to 15

V (nominally 2.14 MV cm$^{-1}$), and durations of either 500 ns or 1 μs. The voltage pulses applied to the thin film capacitor were synchronized to x-ray pulses from specific electron bunches in the SPring-8 storage ring.[10] This experimental arrangement is illustrated in Figure 1(a). A delay generator was used to vary the time between the application of an electric pulse and the arrival of x-rays from one electron bunch, allowing time-resolved measurements of the response of the film to the electric field to be assembled over the course of a series of several million voltage pulses.

In bulk, BFO is a ferroelectric and antiferromagnetic multiferroic with rhombohedral symmetry. The comparison of the structure of BFO with other perovskites is often simplified by describing BFO using a pseudocubic notation in which the pseudocubic lattice parameter is 3.964 Å and the ferroelectric remnant polarization is along the [1 1 1] direction.[11,12] The elastic constraint imposed by the substrate causes BFO thin films to take on a variety of phases depending on the substrate-BFO lattice mismatch and on plastic relaxation phenomena, which in turn depend on the thickness of the BFO layer.[6,13,14]

Multiple structural phases based on distortions of the cubic perovskite unit cell can compete or coexist in highly compressively strained BFO thin films.[6,7,15] The crystallographic symmetry of these phases has been challenging to describe because of the combination of elastic and phase dependent contributions to the structure observed in diffraction experiments. A brief summary of the notation used in other recent studies is given in the supplementary materials. The key distinction, despite this diversity in the notation in the discussion of the phase transformation, is between a (i) rhombohedral-like phase with polarization approximately along the [1 1 1] direction and relatively small out-of-plane lattice parameter and (ii) a more extended tetragonal-like phase with polarization nearly along [0 0 1]. We adopt a notation based on refs. 6 and 7 in which the tetragonal-like phase with out-of-plane lattice parameter $c$=4.64 Å is denoted *T-like* and the rhombohedral-like phase with smaller out of plane lattice parameter $c$=4.18 Å is denoted *R-like*. Other phases are labeled as indicated in Fig. 1.

Figure 1(b) is a diagram of reciprocal space near the (0 0 2) reflection of the LAO substrate, illustrating the reciprocal-space locations of reflections arising from the relevant structural phases of BFO. The approximate four-fold symmetry of the (001) surface of the LAO substrate allows the *R-like* and *tilted-T-like* phases each to have multiple orientations, producing eight variants each.[13] The BFO reflections in Fig. 1(b) thus consist of a single *T-like* phase reflection, eight *tilted-T-like* reflections, and a ring of eight *R-like* reflections centered at a value of the crystallographic index $L$ between the reflections of the *T-like* phase and the LAO substrate. The reciprocal-space location of the distorted *R* phase derived from the bulk BFO structure is not shown in Fig. 1(b). Diffracted intensity arising from the *T-like*, *R-like*, and *tilted-T-like* phases, and from diffuse scattering near the LAO substrate is visible in the steady-state diffraction pattern shown in Figure 1(c). The angular width of x-ray reflections of the BFO layer due to mosaic tilt is sufficiently wide that intensity corresponding to several phases is observed on individual diffraction patterns acquired at a fixed incident angle. Here we focus in detail on the transformation between the *R-like* and *T-like* phases.

**2. Steady-State Structural Transformation**

The transformation between *R-like* and *T-like* phases has been previously induced using electric field pulses provided by scanning probe tips, which results in a steady-state transformation that persists when the applied field is removed.[8] Here we show using synchrotron x-ray diffraction measurements that a similar transformation can also be driven by fields provided using planar thin-film electrodes. The transformation between steady-state configurations that are stable in zero-applied field can be produced using transient electric-field pulses or series of pulses with durations and magnitudes compatible with practical concerns including leakage currents and the finite electrical bandwidth of ferroelectric capacitor structures.

As we discuss further below, the transformation in turn depends on strain developed in the piezoelectric expansion of each phase along the surface normal direction. The piezoelectricity of each phase results in a lattice expansion accompanied by a shift in the Bragg angle of x-ray reflections that can be probed using time-resolved x-ray diffraction. The piezoelectric shift is most apparent for the sharp x-ray reflections of the *T-like* phase, as illustrated in Figure 1(d), which shows rocking curves of the *T-like* (0 0 2) x-ray reflection acquired at times before and during square-wave electric field pulses with a duration of 500 ns and amplitude of 12 V. The 12 V pulses produce a piezoelectric expansion of 0.15% in the applied field. Leakage current through the BiFeO$_3$ thin film results in field lower than the nominal 1.7 MV cm$^{-1}$, by an amount that depends on the pulse history and on the duration and amplitude of the pulses. The A systematic study of the magnitude of the expansion reveals that the piezoelectric shifts are proportional to the magnitude of voltage pulse. The magnitude of the applied electric field, however, is uncertain because the leakage current leads to a resistive voltage drop in the high-sheet resistance LNO bottom electrode that reduces the magnitude of the voltage drop across the BFO capacitor.. The leakage current with 12 V pulses is on the order of 1 mA, with significant variation as a function of time and pulse history. The resistive voltage drop in the bottom electrode corresponding to the maximum leakage current (1 mA) would be 3.3 V, resulting in a reduction of the potential applied to the BFO thin film capacitor to 8.7 V. With the nominal thickness of 70 nm, the range 8.7 to 12 V gives applied electric fields of 1.2 to 1.7 MV/cm. The effective piezoelectric coefficient is thus in the range 10-12 pm/V.

Previous studies have shown that under elastic conditions in which the BFO layer is grown in the *R* phase, applied electric fields yield a piezoelectric distortion with a piezoelectric coefficient of 50 pm/V at low fields.[16] The lower piezoelectric coefficient observed here for the *T-like* phase in comparison with the *R* phase may arise from a low intrinsic piezoelectric coefficient of the *T-like* phase consistent with its very large zero-field *c/a* ratio. Piezoelectric force microscopy (PFM) studies also report smaller piezoelectric coefficients in the *T-like* phase than in the *R* phase.[8]

The transformation between phases results in the redistribution of x-ray intensity in the diffraction patterns of the BFO thin film. Figure 2 shows rocking curves of the *T-like* and *R-like* (0 0 2) reflections in the initial as-grown state, after the application of a series of positive voltage pulses to the top electrode, and finally after the subsequent application of a series of negative voltage pulses. The acquisition of Fig. 2 consisted of (i) an initial diffraction study, (ii) a series of $10^4$ +15 V pulses, (iii) a diffraction study of the state induced by the positive voltage, (iv) a sequence in which two -15 V pulses are followed by a series of $10^3$ pulses with a magnitude of -7 V, and (v) a diffraction study of the state reached following the negative voltage pulses. This pulse sequence was selected arbitrarily to illustrate the difference between diffraction patterns of the structural states reached following positive and negative voltage pulses. The peak intensity of the *T-like* phase reflection increases by a factor of 3 after the positive voltage pulses and returns approximately to its initial intensity after the series of negative voltage pulses, as in Fig. 2(a). The intensities of the *R-like* reflections decrease following positive voltage pulses and subsequently increase after negative voltage pulses as shown in Figs. 2(b) and (c).

The fraction of the film transformed between *R-like* and *T-like* phases can be estimated using the change in the intensity of the x-ray reflections. We make two assumptions in order to simplify this estimation. First, we note that angular range of Figure 2(b) and (c) captures only four of the eight *R-like* reflections. We thus assume that the intensities of the four *R-like* reflections not probed in Figs. 2(b) and (c) vary in the same way as the observed reflections. Combining this assumption with a subtraction of the background underlying the *R-like* phase, we find that the *T-like* phase provides 55% of the total reflected intensity of the BFO film in the initial state, increases to 72% after the positive pulses, and returns to 52% after the negative pulses. The scattering factors of the R-like and T-like phases were estimated using atomic positions from the DFT calculations, which show that the scattering factor of the *R-like* phase is 24% larger than that of the *T-like* phase. A given number of unit cells in the *R-like* phase thus diffract with an intensity that is larger by a factor of 1.54 than the same number of atoms in the *T-like* phase. With these differences in structure factor taken into account, the initial population in the T-like phase is 65%. This observation agrees with previous observations of the relative phase populations in BFO layers produced under similar conditions. Approximately half of the BFO film was in the T-like phase in PFM measurements reported in ref. 8. The observed change in the reflected intensity corresponds to the transformation of 15% of the film from R-like to T-like phases following the applied electric field pulse.

The transformation is reproducible over several cycles of the applied voltage. Figure 3 shows the changes in intensity of the *T-like* phase reflection and two of the *R-like* phase reflections after the application of a sequence of positive and negative voltage pulses. As expected from Fig. 2, the intensity of the *T-like* reflection increases after positive pulses and decreases after negative pulses. The intensity of the *R-like* reflections show the opposite trend as the *T-like* phase, decreasing intensity after positive pulses and increasing after negative pulses. Positive pulses thus lead to a transition from the *R-like* to *T-like* phase, which is reversed by negative voltage pulses. Nominally, large applied voltages of either sign would favor the transformation to the *T-like* state. We hypothesize that the asymmetry between positive and negative voltage pulses arises from effects due to imprint, similar to what has been previously observed in PFM studies.[6]

Voltage pulses of alternating sign result in the redistribution of the structure of the BFO film among the eight structural variants of the *R-like* phase, thus indicating that the reverse transformation favors particular *R-like* variants. As has been previously observed in PFM studies, symmetry breaking resulting from the combination of the direction of the electric field and built-in stresses can result in the selection of a particular variant.[17] X-ray reflections from the *R-like* phase can be denoted by their appearance at relatively high or low values of the x-ray incident angle $\theta$, and their location at the right or left side of detector images of diffraction patterns. The *R-like* (left, low $\theta$) and *R-like* (right, high $\theta$) reflections exhibit a permanent decrease in intensity after completing a cycle of positive and negative pulses. The *R-like* (left, high $\theta$) and *R-like* (right, low $\theta$) reflections exhibit a full recovery or increase in intensity after a cycle of positive and negative voltage pulses. This redistribution of intensity within the family of the *R-like* reflection is indicative of the more favorable selection of one of the structural variants upon the reverse of the phase transition, an effect can arise from elastic constraints imposed on the different domains within the film by epitaxy. Domain patterns for rhombohedral ferroelectrics, for example, are modified by the elastic constraints.[18] The elastic symmetry between *R-like* variants can easily be modified by a small degree of plastic relaxation during epitaxial growth.[19]

## 4. Computational Study of Phase Transformation

While the x-ray diffraction experiments reveal the symmetries of the phases that transform under application of the electric field, understanding the details of this transformation and its energetics requires atomistic simulation of the electronic structure of each BFO phase under comparable strain conditions. To this end, the free energies of BFO layers with different crystallographic symmetries were computationally compared under elastic conditions matching the experiments using DFT[20,21] calculations. The *R-like* and *T-like* phases were modeled using the monoclinic *Cc* and *Cm* space groups, respectively. The true tetragonal *T* phase is modeled using *P*4*mm* space group. The *R-like* phase has both in-plane and out-of-plane polarization components whereas, in the *T* phase, the in-plane polarization is zero.[22,23] The insets of Fig. 4 show the atomic arrangements for of these phases. Further detail regarding the structural models employed in the calculations are given in the supplemental materials.

The DFT calculations were performed using the Vienna Ab initio Simulation Package (VASP)[24,25] with the Perdew-Burke-Ernzerhof (PBE) spin density approximation[26,27] in the generalized gradient (σGGA) scheme. An on-site Coulomb parameter $U$=8 eV was applied to account for the increased Coulomb repulsion between the Fe 3*d* states.[28] The value of the $U$ parameter was chosen using a procedure described in detail below in order to match results of the GGA+$U$ approximation with a more computationally expensive Heyd-Scuseria-Ernzerhof (HSE) functional.

Magnetic moments on Fe atoms in all calculations were arranged in a G-type antiferromagnetic configuration. The projector-augmented plane-wave method,[29,30] using GGA-PBE energies modeled the core-valence electron interactions with pseudovalences of $6s^2 5d^{10} 6p^3$ for Bi, $3p^6 4s^1 3d^7$ for Fe, and $2s^2 2p^4$ for O. A basis set wave-function cutoff of 1000 eV was used to converge the calculated energies of rhombohedral *R*, tetragonal *T*, *R-like*, and *T-like* and cubic perovskite $Pm\bar{3}m$ BFO polymorphs to within 1 meV per formula unit (f.u.). Γ-point centered 7 × 7 × 4 Monkhorst-Pack (MP) *k*-point meshes for the *T*, *T-like* and *R-like* phases (a 7 × 7 × 3 mesh for the *R* phase) sampled the Brillouin zone, converging the energies to within 10 meV/f.u.. Forces on individual ions were relaxed to less than 1 meV/Å and the stress tensor components $\sigma_{ij}$ (with indices *i* and *j* corresponding to x,y,z) on the simulation cell were relaxed to less than 0.1 kbar. The *R* phase was simulated using a primitive cell containing 30 atoms (six f.u.) while the *T*, *T-like*, and *R-like* phases were simulated using $\sqrt{2} \times \sqrt{2} \times 2$ supercells of the pseudocubic unit cell containing 20 atoms (four f.u.). This large supercell containing four formula units of BiFeO$_3$ for the *T, T-like,* and *R-like* phases was required to accommodate the G-type antiferromagnetic configuration. Elastic boundary conditions arising from epitaxy were simulated by fixing the value of in-plane pseudo-cubic lattice constant *a* and relaxing the out-of-plane lattice constant *c*. For the monoclinic phases, the values of *c* and the monoclinic angle $\beta$ were relaxed simultaneously.

The value of *U* in the σGGA+*U* calculations used for all polymorphs described here was optimized to reproduce simultaneously the band gap (2.779 eV) and the local Fe magnetic moment (4.456 $\mu_B$) of the *R* structure obtained using the HSE screened hybrid functional (3.243 eV and 4.119 $\mu_B$).[31,32] We note that this value of *U* differs significantly from other investigations.[33,34] The calculations using the HSE functionals agree with previously published values of band gap (3.4 eV) and local magnetic moment (4.1 $\mu_B$).[35] Fitting *U* to these HSE quantities should mimic the more exact description of exchange contained in HSE without introducing spurious effects potentially created by reproducing experimental data. Due to the computational expense of the HSE calculation, a reduced MP mesh of 6 × 6 × 3 centered at Γ and a plane wave basis cutoff energy of 700 eV were used. For comparison, conducting the σGGA-type calculations with the same Brillouin zone sampling and energy cutoff resulted in a 6% contraction of the volume, a 35% reduction in the band gap and a 2% diminishing of the magnetic moment.

The DFT calculations were first used to estimate the ranges of energetic stability for each these competing phases under various epitaxial strains. The *c*/*a* ratio of the *R-like* phase is always less than that of the *T-like* phase at equal value of *a*. The lower *c*/*a* ratio in the former may result from the stress relaxation achieved through oxygen cage rotations. All of these phases are stable under compressive strain of magnitude $\xi_i$, which is calculated with respect to the pseudo-cubic lattice parameter of the *R* phase. Figure 4(a) plots changes in total energy *E* per f.u. (with the energy of the *R* structure taken as zero) with respect to varying pseudo-cubic in-plane lattice constant *a*. Under all experimentally achievable values of epitaxial strain, the *T-like* phase is energetically more favorable than the *T* phase, in agreement with previous calculations.[22] The *R-like* phase is energetically preferable over *T-like* phase at compressive strains $\xi_i$ less than 3%. The *R-like* phase with $\xi_i=0$ has $c/a=1$ and is therefore equivalent to the *R* phase. Increasing $\xi_i$ to 3% results in crossing of the energies of the *R-like* and *T-like* phases. At this strain the *T-like* phase has a $c/a$ ratio of 1.28. A 4$^{th}$ order polynomial fit to each structure's energies over varying in-plane lattice parameter allows identification of the value of *a* where the *R-like* and *T-like* phases have equal energies. The two fits representing the *R-like* and *T-like* phases intersect at $a_{R-T}$ = 3.881 Å.

The elastic influence of an out-of-plane electric field was approximated by elongating or compressing the pseudo-cubic unit cell in this direction and allowing all the ionic forces to relax to small values. We have not included electric field explicitly in the calculations because the static *U* parameter in tandem with the enhanced electron gas correlation of GGA-PBE needed to open the DFT band gap for $BiFeO_3$ would be unable to describe the changes in electron correlation that an applied electric field would effect. Instead, we rely on the more well-defined ground state inside the σGGA+*U* approximation at varying cell compressions or elongations. Fig. 4(b) shows the variation of the total energy *E* with respect to the value of *c* for the competing monoclinic phases. The minimum of each curve lies at the optimized value of *c* for that phase under the common epitaxial condition *a* = 3.881 Å.

## 5. Phase Transformation Dynamics

A close examination of the DFT results in Fig. 4(b) provides insight into a possible mechanism for the transformation from *R-like* to *T-like* phases. The phase coexistence in the as-deposited film indicates that the energies of the *R-like* and *T-like* phases are approximately equal, consistent with the intersection of the curves showing free-energy as a function of lattice parameter in Fig. 4(a). Piezoelectric expansion in an applied field raises the free energy of both phases, as can be seen in Fig. 4(b) by moving towards larger $c$ along each free energy curve. Note, however, that the curvature of the free energy of the *R-like* phase as a function of $c$ is much higher, so that for any equal fractional expansion of the $c$ axis lattice parameters of the two phases, the free energy of the piezoelectrically expanded *R-like* phase is greater than the expanded *T-like* phase. We thus expect the transformation from *R-like* to *T-like* structures to occur, at least partially, at arbitrarily small values of the applied electric field. The extent of the transformation at each expansion is likely limited by longer-range elastic effects involving the interface between phases, an effect not included in these single-phase DFT calculations. Based on the computational results, we can expect the transformation to become increasingly more favorable at higher applied fields in terms of lowering the free energy. We can also expect that near the boundary between phases, as in the present case, small applied fields will be sufficient to transform some fraction of the material.

In order to investigate the transformation in small applied electric fields experimentally, we now turn from the steady structural state study reported in Figs. 2 and 3 to the dynamics of the phase transformation. In order to probe these dynamics, a series of voltage pulses with duration 500 ns and amplitudes from 4 V to 12 V were applied to the top electrode with a repetition rate of 19 kHz. Diffraction patterns were acquired at a series of delay times following each pulse, and assembled over the course of many pulses to provide sufficient integrated intensity for analysis.

The transient structural response to each set of voltage pulse consists of a rapid transformation of part of the BFO layer from the *R-like* phase to the *T-like* phase, the establishment of a quasi-steady-state coexistence between phases that persists until the electric field is removed, and a transformation back to the zero-field state following the removal of the field. The transformation driven by the applied field during the nanosecond electric field pulses occurs in tandem with the steady-state component probed in Figures 2 and 3. The magnitude of the change in the *T-like* phase intensity is approximately equal to the static intensity changes shown in Figure 2. The stroboscopic methods that were employed to determine the nanosecond transformation behavior probe only the component of the transformation that relaxes to the original configuration between pulses.

We first examine how the extent of the transformation depends on the amplitude of the applied voltage pulses. Figure 5(a) shows the fractional changes in the integrated intensities of the *T-like* and *R-like* x-ray reflections phases during nanosecond voltage pulses with amplitudes from 4 V to 12 V. The measurements in Figure 5(a) are based on diffraction patterns acquired 350 ns after the beginning of the 500 ns voltage pulse, a time at which the transformation has reached steady state conditions in the applied field. The extent of the transformation increases as expected at higher amplitudes: the increase in the intensity of the *T-like* phase is greater at high amplitudes and the decrease of the intensity of the *R-like* phase reflections is also larger. At the largest voltages approximately 10% of the film is transformed from *R-like* phase to *T-like* phase during the pulse.

The dynamics of the phase transformation were probed by monitoring the initial transient in the intensity of the *R-like* phase reflection at the beginning of the voltage pulse. Measurements of the intensity as a function of elapsed time show that the transient occurs with an exponential time dependence, as shown in the supplementary materials. The exponential time constant was independent of the magnitude of the applied voltage and had a value of 90 ns. The relaxation following the end of a 12 V pulse occurs with a time constant of approximately 60 ns. Approximately 96% of the change in the intensity of the R-like phase has recovered in a period of 150 ns after the end of the electric field pulse, as in supplementary Fig. S3. The near equivalence of the time constants for the transformation and the reverse transformation, the lack of dependence of the charging time constant on the magnitude of the voltage pulse, and the similarity of their absolute values to the characteristic charging time of the device suggest that the time dependence observed here is limited by the slow time-variation of the electric field pulse rather than by the dynamics of the transformation. The 90 ns time constant for the transformation should thus be taken as an upper limit for the actual rate of the transformation.

The transient transformation during applied electric fields occurs differently in the different structural variants of the *R-like* phase. Fig. 5(b) shows the change in the peak intensities of the four different *R-like* phase reflections that fell within the range of the angular scans. The *R-like* phase reflections are labeled in Fig. 5(b) with their angular location, either right or left of the diffraction pattern and at low or high values of the Bragg angle $\theta$. The two *R-like* reflections appearing on the left side of the diffraction pattern, *R-like* (left, low $\theta$) and *R-like* (left, high $\theta$) exhibit large changes in intensity in applied fields, decreasing in intensity by 30% during voltage pulses with amplitude of 12 V. The two reflections on the right side of the diffraction pattern exhibit a far smaller change of approximately 5% in the same field. Nearly all of the transient transformation during applied voltages is the result of material transforming from the *R-like* (left) variants to the *T-like* phase. The observation that the *R-like* (left) variant transforms more readily suggests the switching dynamics depend on the stress remaining from epitaxial growth or on a similar artifact breaking the symmetry between structural variants. The difference between *R-like* structural variants was observed in all of the capacitors probed.

The transient measurements show that the phase transformation between *R-like* and *T-like* phases is highly reproducible and changes negligibly over the many millions of pulses required to assemble the stroboscopic measurements. In the particular case of Fig. 5, the dynamical studies required the application of $2.1 \times 10^8$ voltage pulses. In part this high degree of reversibility may arise from the small free-energy difference between *R-like* and *T-like* phases, which results in low energy dissipation during both the forward and reverse transformations.

Several conclusions can be drawn from the observation of both transient reversible change of phases and long-term quasi-stable transformation between phases. First, the existence of the quasi-stable transformation by individual pulses (e.g. as exhibited in Figs. 2 and 3) suggests that the local energy landscape is complex and that hysteresis effects can allow either state to have the minimum energy configuration. The BFO thin film can, under the influence of the applied field, be driven away from the free energy minimum and can transform during the field into the phase favored by the free-energy calculation shown in Fig. 4. The rapid reversibility of the transient transformation following the removal of the applied electric field shows that there is a minimal barrier for the transformation. The reproducibility over large number of cycles suggests that there is minimal energy dissipation during the transient component of the transformation. Finally, we note that it is likely that at least some of the processes resulting in the quasi-steady-state changes apparent in Figs. 2 and 3 are can in principle be accompanied by relaxation events that occur at times longer than the time required for the experimental study and that there are possible partial relaxation phenomena occurring even at times of hours or longer following the transformation.

## 5. Conclusion

Taken together, the steady-state and dynamical methods suggest a physical picture in which applied electric fields rapidly transform BFO between the *R-like* and *T-like* phases. The rate of this transformation is sufficiently fast that large fractions of the thin film can be transformed within less than 100 ns, and the rate is likely limited by electrical design in this experiment rather than by a fundamental time dependence of the transition. After the pulses, steady-state measurements show that some fraction of the volume of BFO is stable in the transformed state while a separate fraction of the BFO layer returns to the original configuration. The transformation can be driven by small-magnitude (4 to 12 V) applied voltage pulses, which is consistent with computational models of the dependence of the energies of the relevant structural phases on applied strains. The large volume fraction of material associated with the transient transformation as well its reproducibility over many cycles are both favorable for possible applications of the BFO phase transformation both in tunable electronic components and actuators. The incomplete transformation of the thin film as a whole remains an important challenge in adapting the BFO transformation to potential technological applications.

A large compressive stress is necessary to place BFO sufficiently close to the phase boundary to yield the transformation in experimentally realizable fields. BFO grown on substrates providing smaller compressive stress suffers dielectric breakdown before the phase transformation is observed.[16] We note, however, that adjusting the composition of BFO to place it near the phase boundary has recently become possible and that this approach may present similarly favorable transformation dynamics in a more widely applicable form.[36,37]

**Supporting Information**
Supporting Information is available from the Wiley Online Library or from the author.

**Acknowledgements**

MPC, PC, and PGE gratefully acknowledge support by the U. S. National Science Foundation, through Grant No. DMR-1106050. The SPring-8 measurement was supported by the JASRI under Proposal Nos. 2012A1105 and 2012B1551. HJL, SSL, and JYJ acknowledge support by the National Research Foundation (NRF-2014R1A1A3053111 and NRF-2011-220-C00016). LC acknowledges supports by Hong Kong, Macao & Taiwan Science & Technology Cooperation Program of China (#2015DFH10200) and NSFC #11474146. KCP and SMN are thankful to the National Science Foundation for partial support under Grant No. DMR-1309114.

[1] C. H. Ahn, A. Bhattacharya, M. Di Ventra, J. N. Eckstein, C. D. Frisbie, M. E. Gershenson, A. M. Goldman, I. H. Inoue, J. Mannhart, A. J. Millis, A. F. Morpurgo, D. Natelson, J. M. Triscone, *Rev. Mod. Phys.* **2006**, *78*, 1185.

[2] S. E. Park, T. R. Shrout, *J. Appl. Phys.* **1997**, *82*, 1804.

[3] H. Y. Hwang, S.-W. Cheong, N. P. Ong, B. Batlogg, *Phys. Rev. Lett.* **1996**, *77*, 2041.

[4] R. J. Zeches, M. D. Rossell, J. X. Zhang, A. J. Hatt, Q. He, C. H. Yang, A. Kumar, C. H. Wang, A. Melville, C. Adamo, G. Sheng, Y. H. Chu, J. F. Ihlefeld, R. Erni, C. Ederer, V. Gopalan, L. Q. Chen, D. G. Schlom, N. A. Spaldin, L. W. Martin, R. Ramesh, *Science* **2009**, *326*, 977.

[5] H. M. Christen, J. H. Nam, H. S. Kim, A. J. Hatt, N. A. Spaldin, *Phys. Rev. B* **2011,** *83*, 144017.

[6] Z. Chen, S. Prosandeev, Z. L. Luo, W. Ren, Y. Qi, C. W. Huang, L. You, C. Gao, I. A. Cornev, T. Wu, J. Wang, P. Yang, T. Sritharan, L. Bellaiche, L. Chen, *Phys. Rev. B* **2011**, *84*, 094116.

[7] A. R. Damodaran, C. W. Liang, Q. He, C. Y. Peng, L. Chang, Y. H. Chu, L. W. Martin, *Adv. Mater.* **2011**, *23*, 3170.

[8] J. X. Zhang, B. Xiang, Q. He, J. Seidel, R. J. Zeches, P. Yu, S. Y. Yang, C. H. Wang, Y-H. Chu, L. W. Martin, A. M. Minor, R. Ramesh, *Nature Nanotechnol.* **2011**, *6*, 98.

[9] O. Sakata, Y. Furukawa, S. Goto, T. Mochizuki, T. Uraga, K. Takeshita, H. Ohashi, T. Ohata, T. Matsushita, S. Takahashi, H. Tajiri, T. Ishikawa, M. Nakamura, M. Ito, K. Sumitani, T. Takahashi, T. Shimura, A. Saito, M. Takahashi, *Surf. Rev. Lett.* **2003**, *10*, 543.

[10] A. Grigoriev, D.-H. Do, P. G. Evans, B. W. Adams, E. Landahl, E. M. Dufresne, *Rev. Sci. Instrum.* **2007**, *78*, 023105.

[11] D. Lebeugle, D. Colson, A. Forget, M. Viret, *Appl. Phys. Lett.* **2007**, *91*, 022907.

[12] O. Diéguez, P. Aguado-Puente, J. Junquera, J. Íñiguez, *Phys. Rev. B* **2013**, *87*, 024102.

[13] Z. Chen, Z. Luo, Y. Qi, P. Yang, S. Wu, C. Huang, T. Wu, J. Wang, C. Gao, T. Sritharan, L. Chen, *Appl. Phys. Lett.* **2010**, *97*, 242903.

[14] Z. Chen, Z. Luo, C. Huang, Y. Qi, P. Yang, L. You, C. Hu, T. Wu, J. Wang, C. Gao, T. Sritharan, L. Chen, *Adv. Funct. Matter.* **2011**, *27*, 133.

[15] A. Hatt, N. Spaldin, C. Ederer, *Phys. Rev. B* **2010**, *81*, 054109.

[16] P. Chen, R. Sichel-Tissot, J. Y. Jo, R. T. Smith, S. H. Baek, W. Saenrang, C. B. Eom, O. Sakata, E. M. Dufresne, P. G. Evans, *Appl. Phys. Lett.* **2012**, *100*, 062906.

[17] L. You, Z. Chen, X. Zou, H. Ding, W. Chen, L. Chen, G. Yuan, J. Wang, *ACS Nano*, **2012**, *6*, 5388.

[18] A. E. Romanov, M. J. Lefevre, J. S. Speck, W. Pompe, S. K. Streiffer, and C. M. Foster, *J. Appl. Phys.* **1998**, *83*, 2754.

[19] R. J. Sichel, A. Grigoriev, D.-H. Do, S.-H. Baek, H.-W. Jang, C. M. Folkman, C.-B. Eom, Z. Cai, P. G. Evans, *Appl. Phys. Lett.* **2010**, *96*, 051901.

[20] P. Hohenberg, W. Kohn, *Phys. Rev.* **1964**, *136*, B864.

[21] W. Kohn, L. J. Sham, *Phys. Rev.* **1965**, *140*, A1133.

[22] H. Béa, B. Dupé, S. Fusil, R. Mattana, E. Jacquet, B. Warot-Fonrose, F. Wilhelm, A. Rogalev, S. Petit, V. Cros, A. Anane, F. Petroff, K. Bouzehouane, G. Geneste, B. Dkhil, S.

Lisenkov, I. Ponomareva, L. Bellaiche, M. Bibes, A. Barthélémy, *Phys. Rev. Lett.* **2009**, *102*, 217603.

[23] H.-J. Liu, C.-W. Liang, W.-I. Liang, H.-J. Chen, J.-C. Yang, C.-Y. Peng, G.-F. Wang, F.-N. Chu, Y.-C. Chen, H.-Y. Lee, L. Chang, S.-J. Lin, Y.-H. Chu, *Phys. Rev. B* **2012**, *85*, 014104.

[24] G. Kresse, J. Furthmüller, *Phys. Rev. B* **1996**, *54*, 11169.

[25] G. Kresse, J. Furthmüller, *Comp. Mater. Sci.* **1996**, *6*, 15.

[26] J. P. Perdew, K. Burke, M. Ernzerhof, *Phys. Rev. Lett.* **1996**, *77*, 3865.

[27] J. P. Perdew, K. Burke, M. Ernzerhof, *Phys. Rev. Lett.* **1997**, *78*, 1396.

[28] V. I. Anisimov, F. Aryasetiawan, A. Lichtenstein, *J. Phys.:Condens. Matter* **1997**, *9*, 767.

[29] P. E. Blöchl, *Phys. Rev. B* **1994**, *50*, 17953.

[30] G. Kresse, D. Joubert, *Phys. Rev. B* **1999**, *59*, 1758.

[31] J. Heyd, G. E. Scuseria, M. Ernzerhof, *J. Chem. Phys.* **2003**, *118*, 8207.

[32] J. Heyd, G. E. Scuseria, M. Ernzerhof, *J. Chem. Phys.* **2006**, *124*, 219906.

[33] J. Neaton, C. Ederer, U. Waghmare, N. Spaldin, K. Rabe, *Phys. Rev. B* **2005**, *71*, 014113.

[34] H. Tutuncu, G. Srivastava, *J. Appl. Phys.* **2008**, *103*, 083712.

[35] A. Stroppa, S. Picozzi, *Phys. Chem. Chem. Phys.* **2010**, *12*, 5405.

[36] H. Hojo, K. Onuma, Y. Ikuhara, M. Azuma, *Appl. Phys. Exp.* **2014**, *7*, 091501.

[37] Z. Fan, J. X. Xiao, H. J. Liu, P. Yang, Q. Q. Ke, W. Ji, K. Yao, K. P. Ong, K. Y. Zeng, J. Wang *ACS Appl. Mater. Interfaces* **2015**, *7*, 2648.

**Figure 1.** (a) Experimental arrangement for time-resolved x-ray microdiffraction experiments. The BiFeO$_3$ thin film is composed of distinct blocks with differing crystallographic symmetry. (b) Reciprocal-space locations of x-ray reflections arising from BiFeO$_3$ and LAO near the LAO substrate (002) reflection. (c) X-ray diffraction pattern acquired with photon energy 12.4 keV at Bragg angle $\theta$=13.84°, near the Bragg condition for the *R-like* reflections. The powder x-ray diffraction ring arising from the Au top electrode is indicated by (*). (d) Piezoelectric distortion of the BFO *T-like* phase during a 500 ns-duration square-waveform voltage pulse with a magnitude of 12 V, corresponding to a nominal electric field of $1.7 \times 10^6$ V cm$^{-1}$. The piezoelectric expansion of the *T-like* phase in this applied field is 0.15%.

**Figure 2.** Rocking curves of the (a) *T-like*, (b) *R-like* (left), and (c) *R-like* (right) x-ray reflections of BFO before the application of an electric field (black), after a series of $10^4$ repetitions of a +15 V triangular voltage pulse with a 1 µs durations applied to the top electrode (red), and subsequently applying a series of negative pulses consisting of two repetitions of a pulse with magnitude -15 V and $10^3$ repetitions of pulses with a magnitude of -7 V (green). Diffraction patterns (inset) are acquired near the Bragg conditions of the (i) *T-like* and (ii) *R-like* reflections. Black rectangles in the inset illustrate the regions of interest over which the diffracted intensity was integrated at each point in the rocking curve scans. Diffraction from the top electrodes of the capacitors is indicated with (*) in the inset. The splitting of the *T-like* x-ray reflection in (a) arises from the twinning of the LAO substrate. The maxima at low and high angles in (b) and (c) are termed (left) and (right) in the text, respectively.

**Figure 3.** Normalized change in the peak intensities of x-ray reflections from *T-like* and *R-like* structural phases following voltage pulses in a sequence of different polarities and magnitudes. The intensity of the *T-like* phase reflection increases after the application of positive voltage pulses (shaded) and decreases after negative pulses (unshaded), while the *R-like* phase intensities show the opposite trend. The regions of interest over which the intensity is integrated are indicated in the inset in Figure 2.

**Figure 4.** Total energies per formula unit of BFO phases relative to the relaxed bulk *R* structure as a function of lattice parameter under elastic conditions corresponding to (a) epitaxial growth and (b) subsequent elastic deformation via the converse piezoelectric effect. (a) Energies of the *R-like* (modeled as *Cc* symmetry), *T-like* (*Cm*) and *T* (*P4mm*) phases as a function of the in-plane lattice parameter *a*, where the out-of-plane *c* lattice parameter and monoclinic angle *β* are fully relaxed at each value of *a*. (b) Energies of the *R-like* and *T-like* phases as a function of the pseudo-cubic out-of-plane lattice parameter *c*, at fixed *a*=3.881 Å. Model structures of the *T*, *T-like* and *R-like* phases are inset in (a) and (b). Green arrows indicate spin directions on Fe ions.

**Figure 5.** (a) Change of the integrated intensity during 500 ns-duration voltage pulses of the *T-like* phase reflection and the total integrated intensity of the four *R-like* phase reflections accessible in the angular range of the x-ray rocking curves shown in Fig. 2. (b) Change in peak intensities of four individual *R-like* reflections.

Cosgriff, *et al*., Figure 1

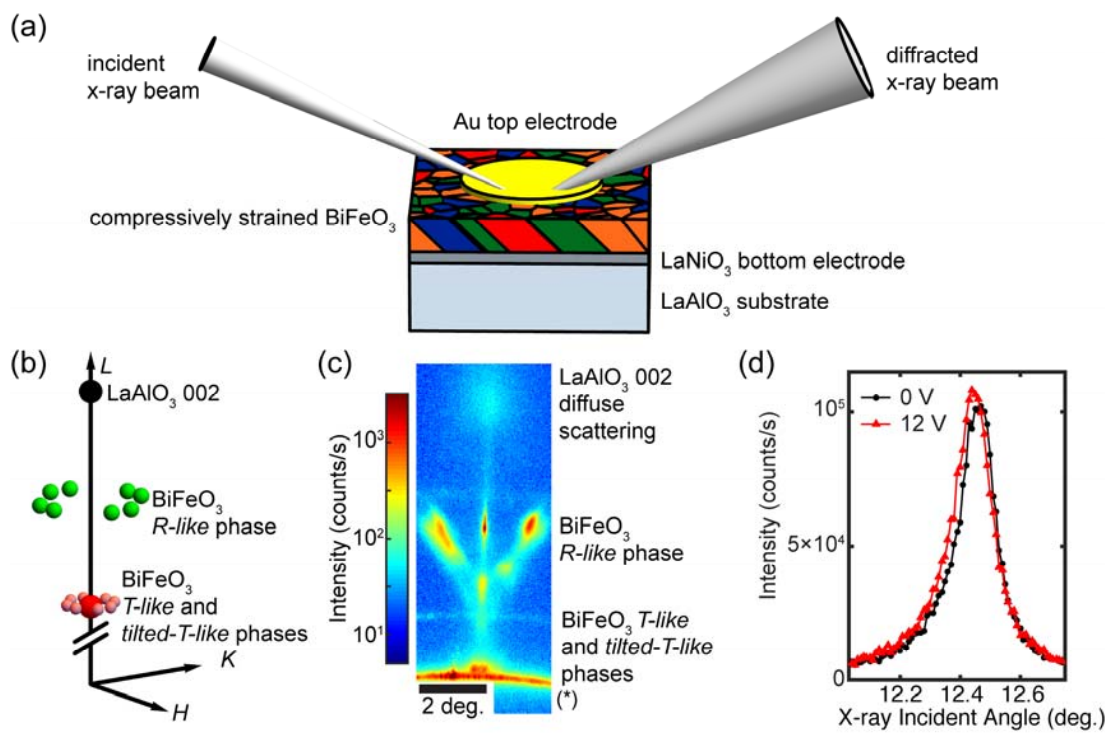

Cosgriff, *et al*., Figure 2

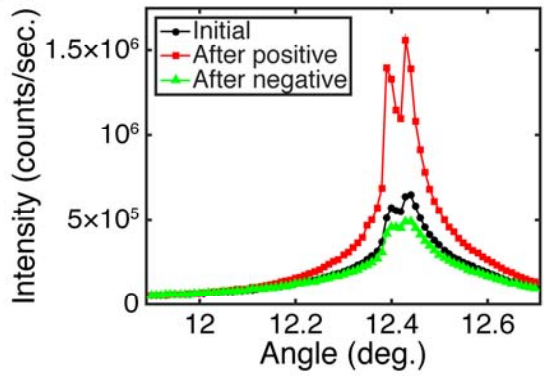
(a) *T-like*

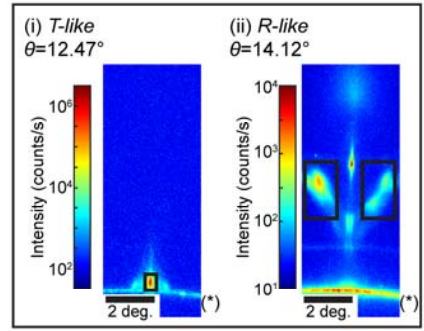
(i) *T-like* θ=12.47°
(ii) *R-like* θ=14.12°

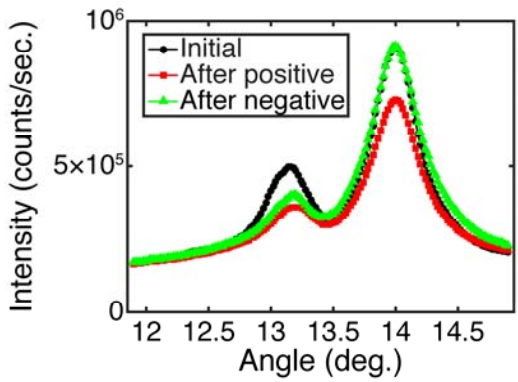
(b) *R-like* (left)

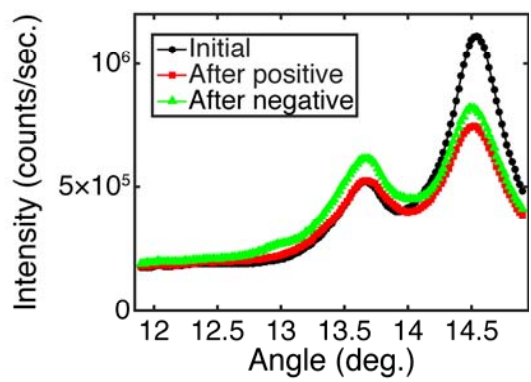
(c) *R-like* (right)

Cosgriff, *et al*., Figure 3

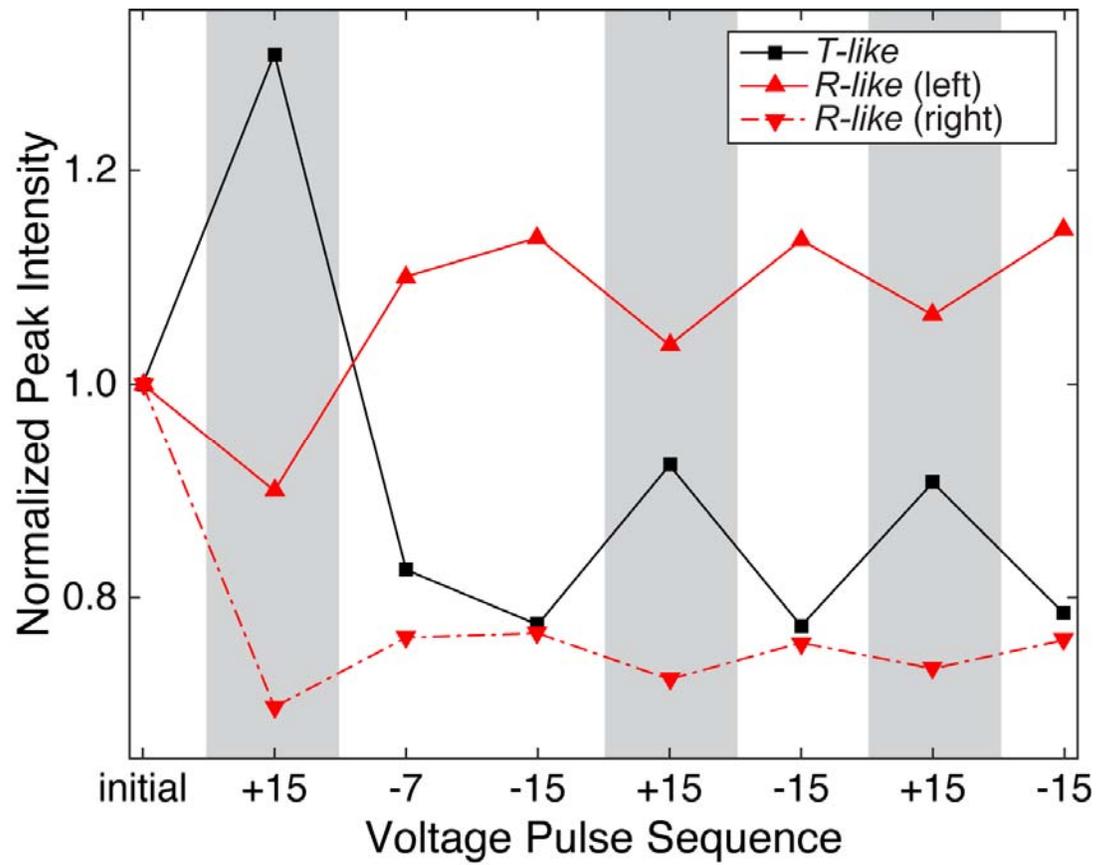

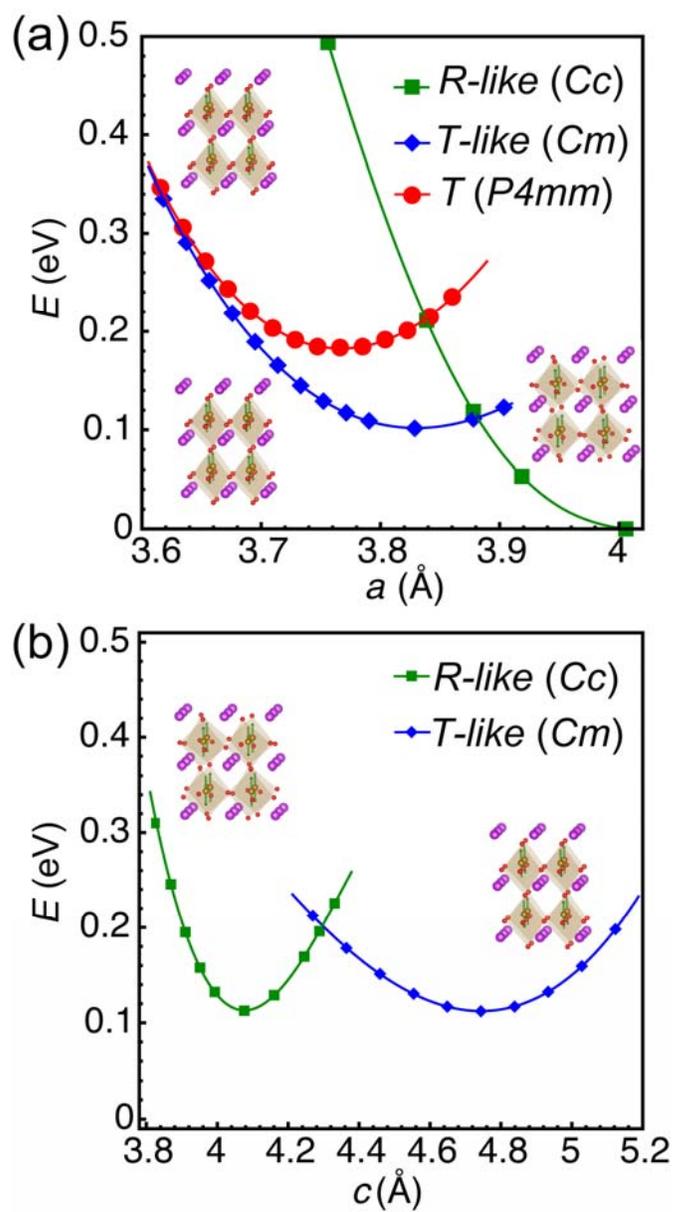

Cosgriff, *et al*., Figure 5

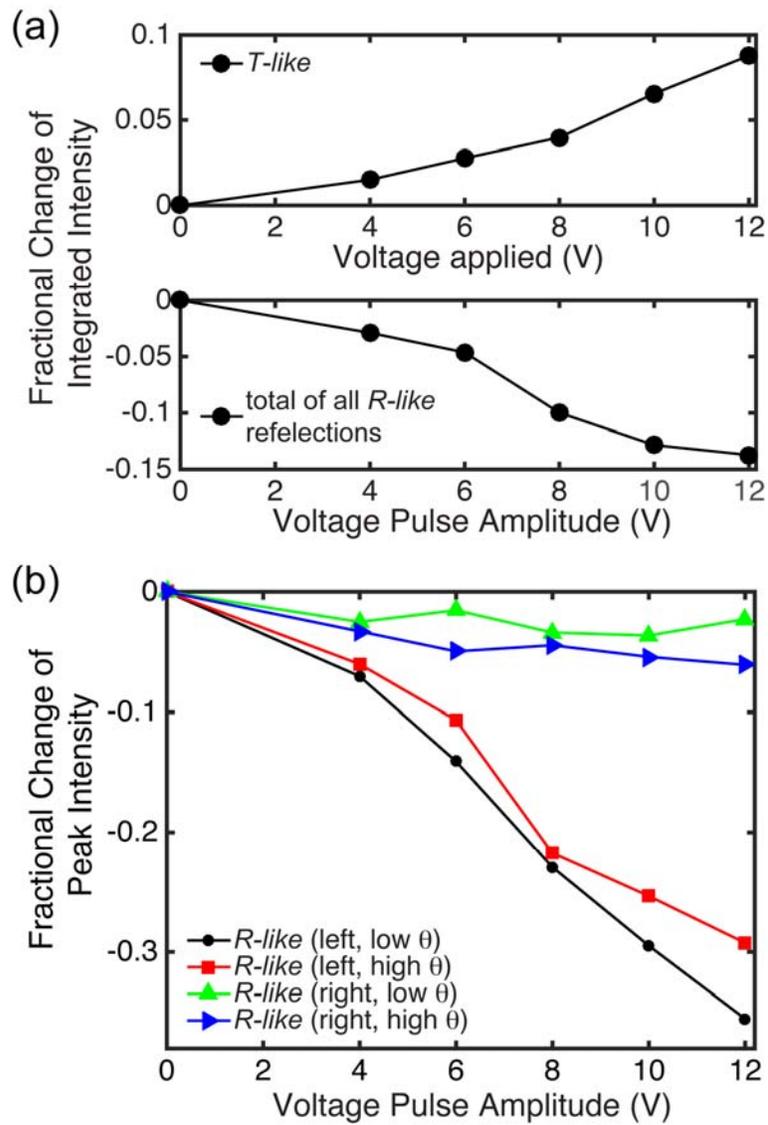

# Supporting Information

**Nanosecond Phase Transition Dynamics in Compressively Strained Epitaxial BiFeO$_3$**

*Margaret P. Cosgriff, Pice Chen, Sung Su Lee, Hyeon Jun Lee, Lukasz Kuna, Krishna C. Pitike, Lydie Louis, William D. Parker, Hiroo Tajiri, Serge M. Nakhmanson, Ji Young Jo, Zuhuang Chen, Lang Chen, and Paul G. Evans\**

Dr. M. P. Cosgriff, Dr. Pice Chen,[+] Prof. P. G. Evans,
Department of Materials Science and Engineering and Materials Science Program, University of Wisconsin-Madison, Madison, WI 53706, USA
E-mail: pgevans@wisc.edu

S. S. Lee, H. J. Lee, Prof. J. Y. Jo
Gwangju Institute of Science and Technology, Gwangju 500712, South Korea

L. Kuna,
Department of Physics, University of Connecticut, Storrs, Connecticut 06269, USA

K. C. Pitike, Dr. L. Louis,
Department of Materials Science and Engineering, and Institute of Materials Science, University of Connecticut, Storrs, Connecticut 06269, USA

Prof. S. M. Nakhmanson,
Department of Physics, Department of Materials Science and Engineering, and Institute of Materials Science, University of Connecticut, Storrs, Connecticut 06269, USA

Dr. W. D. Parker,
Argonne Leadership Computing Facility, Argonne National Laboratory, Argonne, Illinois 60439, USA and School of Chemical Engineering, Purdue University, West Lafayette, Indiana 47907, USA

Dr. H. Tajiri,
Japan Synchrotron Radiation Research Institute, SPring-8, Hyogo 6795198, Japan

Dr. Z. Chen,
Department of Materials Science and Engineering, University of California, Berkeley, California 94720, USA

Prof. L. Chen,
Department of Physics, University of Science and Technology of China, Shenzhen, China 518055

[+] Present address: Department of Materials Science and Engineering, Northwestern University, Evanston, Illinois 60208, USA

## 1. Piezoelectricity of *T-like* Phase

The piezoelectricity of the *T-like* phase of BFO was probed using square-wave voltage pulses with amplitudes ranging from 4 V to 12 V. A rocking curve scan of the Bragg angle $\theta$ was acquired for each magnitude of the voltage. The piezoelectric expansion $\varepsilon$ was determined from the angular peak shift $\Delta\theta$ induced by the electric field using $\varepsilon = \cot(\theta)\,\Delta\theta$, as shown in Figure S1. The piezoelectric coefficient $d_{33}$ was estimated from the structural data by using the nominal BFO film thickness of 70 nm to determine the electric field $E$ and fitting the data with $\varepsilon = d_{33} E$. The best fit of this piezoelectric model is shown as the solid line in Figure S1, which gives $d_{33} = 9.3$ pm/V. We note that there was significant leakage current during the voltage pulses and that there is thus a significant uncertainty in the voltage drop across the BFO thin film capacitors. The value of $d_{33}$ determined from Figure S1 is thus a lower limit for the actual piezoelectric coefficient. Based on the measured leakage currents and bottom electrode resistance, up to 50% of the applied voltage can fall across the bottom electrode rather than across the thin-film capacitor sample.

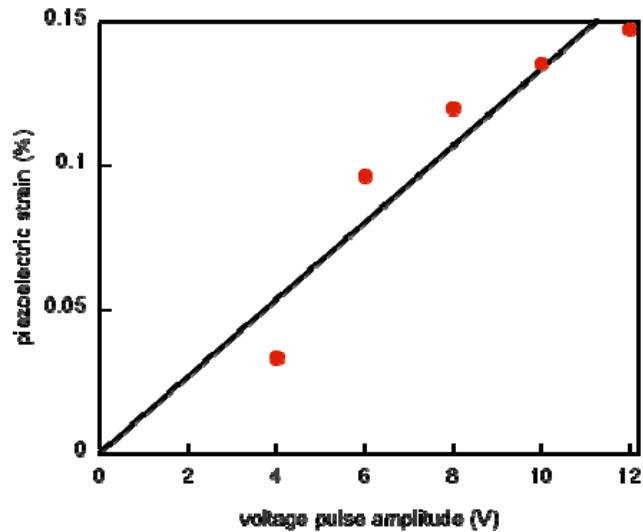

**Figure S1.** Piezoelectric expansion of the *T-like* phase of BFO (circles) as a function of the amplitude of applied voltage pulses. The line is fit with a linear piezoelectric expansion model.

## 2. Structural Models for *R-like* and *T-like* Phases, Phase Notation

Models of the structural phases used for the simulation are shown in Fig. S2. Key differences between the structures of the two phases are the ratio of lattice parameters along the in-plane and out-of plate directions, labeled *a* and *c* in Fig. S2(a), and the rotation of the oxygen octahedra. A summary of the notation used to label structural phases of compressively strained BFO in this work and several recent reports is given in Table S1. Table S2 reports DFT values calculated for three symmetries of BiFeO$_3$ comparing three separate approximations on the exchange-correlation functional.

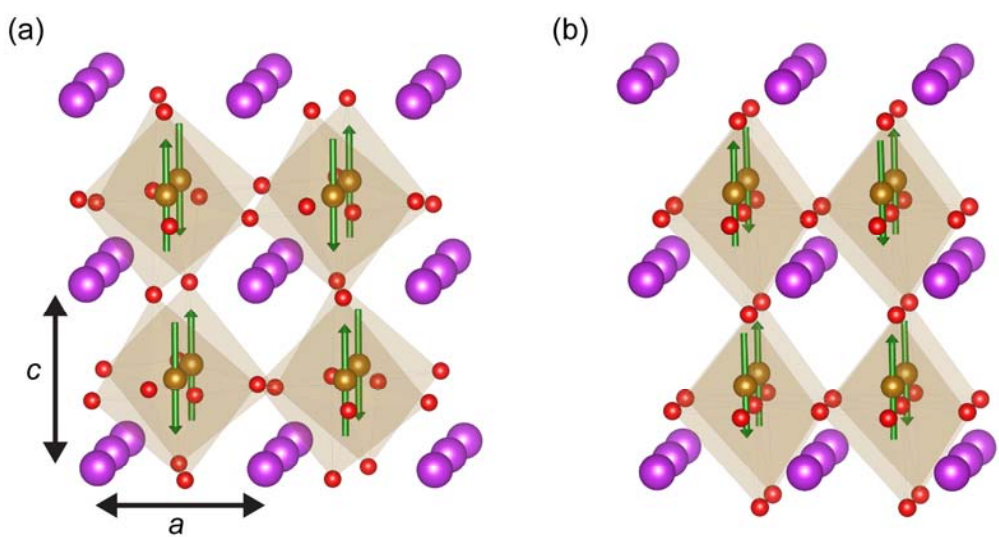

**Figure S2.** Structures of the (a) *R-like* (*Cc* symmetry) rhombohedral-like phase and (b) *T-like* (*Cm* symmetry) tetragonal-like phases considered in DFT calculations.

**Table S1.** Notation describing structural phases in BFO thin films on LAO substrates.

| 1. This Work | | | | | |
|---|---|---|---|---|---|
| Phase Notation | R-like | T-like | Tilted T-like | R | T |
| Assumed Symmetry for Calculations | Cc | Cm | | R3c | P4mm |
| **2. Previous Reports** | | | | | |
| Chen 2011[S1] | Tri-1 | T-like Mc | Tri-2 | R-like $M_A$ | |
| Damodaran 2011[S2] | $M_I$ | $M_{II}$ | $M_{II}$-tilt | R | |
| He 2011[S3] | R | T | | | |
| Christen 2011[S4] | | T-like Mc | | | |
| You 2012[S5] | $R_{tilted}$ | T-like | $T_{tilted}$ | | |
| Liu 2012[S6] | $M_I$ | $M_{II}$ | $M_{II}$-tilt | R | |
| Pailloux 2014[S7] | R-like, R | T-like, T | | | |
| Luo 2014[S8] | Tri-1 | T-like | Tri-2 | | |
| Cheng 2015[S9] | $M_I$ | $M_{II}$ | $M_{II}$-tilt | | |

**Table S2.** DFT parameters for three high-symmetry structures of BiFeO$_3$. (*) Indicates that the strain free R-like phase is identical to the bulk R phase in this structural model.

| | R3c (Pure R*) | | | P4mm (Pure T) | | Cm (T-like) | |
|---|---|---|---|---|---|---|---|
| | GGA | GGA+U | HSE | GGA | GGA+U | GGA | GGA+U |
| **Formula units per cell** | 6 | 6 | 6 | 4 | 4 | 4 | 4 |
| **a (Å)** | 5.63 | 5.63 | 5.50 | 5.29 | 5.32 | 5.38 | 5.44 |
| **b (Å)** | 5.63 | 5.63 | 5.50 | 5.29 | 5.32 | 5.34 | 5.39 |
| **c (Å)** | 14.04 | 14.07 | 13.25 | 9.76 | 9.64 | 9.81 | 9.68 |
| **γ (°)** | 120.00 | 120.00 | 120.00 | 90.00 | 90.00 | 85.94 | 86.16 |
| **Volume of unit cell (Å$^3$)** | 385.13 | 385.56 | 346.81 | 273.39 | 274.44 | 281.44 | 283.17 |
| **Magnetic moment (μB)** | 3.72 | 4.46 | 4.11 | 3.75 | 4.45 | 3.74 | 4.45 |
| **Band gap (eV)** | 1.06 | 2.78 | 3.16 | 0.69 | 1.78 | 0.85 | 2.32 |

## 4. Phase Transition Dynamics

The intensities x-ray reflections from the *R-like* phases decreases during voltage pulses due to the transformation of parts of the BFO thin film from the *R-like* phase to the *T-like* phase. The transient response during a voltage pulse with a magnitude of 12 V and duration 500 ns is shown in Figure S3. The transformation begins at the start of the pulse at 0 ns, reaches a steady state after approximately 100 ns, and subsequently relaxes following the end of the pulse at 500 ns. Figure S3 shows the intensity variation for the x-ray reflections terms *R-like* (left) in the text, for which the variation in intensity is larger than the *R-like* (right) reflections.

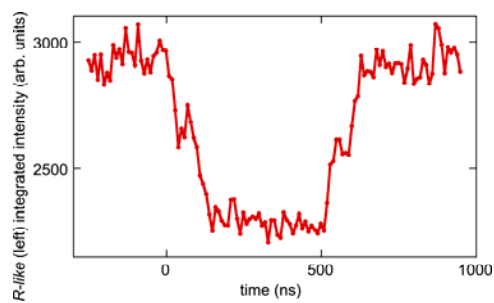

**Figure S3.** Integrated intensity of the *R-like* phase reflections appearing on the left side of the diffraction pattern, termed *R-like* (left) in the main text, as a function of time. A 500 ns applied voltage pulse with amplitude 12 V begins at 0 ns.

The variation of the *R-like* phase intensity following the start of voltage pulses with amplitudes from 4 V to 12 V is shown in Fig. S4(a). The magnitude of the step change in *R-like* phase intensity increases for larger voltage pulse amplitudes. The normalized change in intensity following the beginning of the voltage pulses, Fig. S4(b), can be fit with a single exponential decay with a time constant of 90 ns.

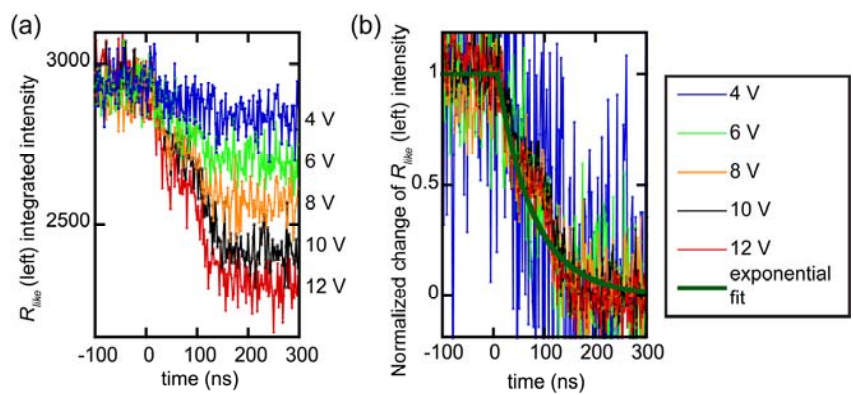

**Figure S4.** (a) Integrated intensity of the *R-like* phase reflections appearing on the left side of the diffraction pattern, termed *R-like* (left) in the main text, as a function of time following the beginning of 500 ns voltage pulses with amplitudes ranging from 4 V to 12 V. (b) Normalized change in *R-like* phase intensity for all voltages. The solid line is an exponential decay with a characteristic time constant of 90 ns.

**Supplemental References**

[S1] Z. Chen, S. Prosandeev, Z. L. Luo, W. Ren, Y. Qi, C. W. Huang, L. You, C. Gao, I. A. Cornev, T. Wu, J. Wang, P. Yang, T. Sritharan, L. Bellaiche, L. Chen, *Phys. Rev. B* **2011**, *84*, 094116.

[S2] A. R. Damodaran, C. W. Liang, Q. He, C. Y. Peng, L. Chang, Y. H. Chu, L. W. Martin, *Adv. Mater.* **2011**, *23*, 3170.

[S3] Q. He, Y.-H. Chu, J. T. Heron, S. Y. Yang, W. I. Liang, C. Y. Kuo, H. J. Lin, P. Yu, C. W. Liang, R. J. Zeches, W. C. Kuo, J. Y. Juang, C. T. Chen, E. Arenholz, A. Scholl, R. Ramesh, *Nature Comm.* **2011**, *2*, 225 (2011).

[S4] H. M. Christen, J. H. Nam, H. S. Kim, A. J. Hatt, N. A. Spaldin, *Phys. Rev. B* **2011,** *83*, 144017.

[S5] L. You, Z. Chen, X. Zou, H. Ding, W. Chen, L. Chen, G. Yuan, J. Wang, *ACS Nano*, **2012**, *6*, 5388.

[S6] H.-J. Liu, C.-W. Liang, W.-I. Liang, H.-J. Chen, J.-C. Yang, C.-Y. Peng, G.-F. Wang, F.-N. Chu, Y.-C. Chen, H.-Y. Lee, L. Chang, S.-J. Lin, Y.-H. Chu, *Phys. Rev. B* **2012**, *85*, 014104.

[S7] F. Pailloux, M. Couillard, S. Fusil, F. Bruno, W. Saidi, V. Garcia, C. Carrétéro, E. Jacquet, M. Bibes, A. Barthélémy, G. A. Botton, J. Pacaud, *Phys. Rev. B* **2014**, *89*, 104106.

[S8] Z. L. Luo, H. Huang, H. Zhou, Z. H. Chen, Y. Yang, L. Wu, C. Zhu, H. Wang, M. Yang, S. Hu, H. Wen, X. Zhang, Z. Zhang, L. Chen, D. D. Fong, C. Gao, *Appl. Phys. Lett.* **2014**, *104*, 182901.

[S9] C.-E. Cheng, H.-J. Liu, F. Dinelli, Y.-C. Chen, C.-S. Chang, F. S.-S. Chien, Y.-H. Chu, *Scientific Rep.* **2015**, *5*, 8091.